\documentclass{WileyMSP-template}

\usepackage{chemformula} 
\usepackage[T1]{fontenc} 
\usepackage[version=3]{mhchem} 

\usepackage{hyperref}
\usepackage{graphicx}
\usepackage{multirow}
\usepackage{titlesec}
\usepackage{amsmath}
\usepackage{multicol}
\newcommand{\rom}[1]{\lowercase\expandafter{\romannumeral #1\relax}}
\titleformat{\section}
  {\bfseries\Large}{\thesection.}{0.5em}{}

\begin{document}

\pagestyle{fancy}
\rhead{\includegraphics[width=2.5cm]{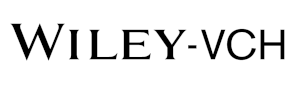}}

\title{Neural Network-Driven Molecular Insights into Alkaline Wet Etching of GaN: Toward Atomistic Precision in Nanostructure Fabrication}

\maketitle


\author{Purun-hanul Kim}

\author{Jeong Min Choi}

\author{Seungwu Han*}

\author{Youngho Kang*}

\dedication{}

\begin{affiliations}
Purun-hanul Kim, Jeong Min Choi, \\
Department of Materials Science and Engineering and Research Institute of Advanced Materials, Seoul National University, Seoul 08826, Republic of Korea \\

Prof. Seungwu Han \\
Department of Materials Science and Engineering and Research Institute of Advanced Materials, Seoul National University, Seoul 08826, Republic of Korea \\
Korea Institute for Advanced Study, Seoul 02455, Republic of Korea \\
Email Address: hansw@snu.ac.kr \\

Prof. Youngho Kang \\
Department of Materials Science and Engineering, Incheon National University, Incheon 22012, Republic of Korea \\
Email Address: youngho84@inu.ac.kr \\

\end{affiliations}



\begin{abstract}

We present large-scale molecular dynamics (MD) simulations based on a machine-learning interatomic potential to investigate the wet etching behavior of various GaN facets in alkaline solution—a process critical to the fabrication of nitride-based semiconductor devices. A Behler-Parrinello-type neural network potential (NNP) was developed by training on extensive DFT datasets and iteratively refined to capture chemical reactions between GaN and KOH. To simulate the wet etching of GaN, we perform NNP-MD simulations using the temperature-accelerated dynamics approach, which accurately reproduces the experimentally observed structural modification of a GaN nanorod during alkaline etching. The etching simulations reveal surface-specific morphological evolutions: pyramidal etch pits emerge on the $-c$ plane, while truncated pyramidal pits form on the $+c$ surface. The non-polar $m$ and $a$ surfaces exhibit lateral etch progression, maintaining planar morphologies. Analysis of MD trajectories identifies key surface reactions governing the etching mechanisms. To gain deeper insights into the etching kinetics, we conduct enhanced-sampling MD simulations and construct free-energy profiles for Ga dissolution, a process that critically influences the overall etching rate. The $-c$, $a$, and $m$ planes exhibit moderate activation barriers, indicating the feasibility of alkaline wet etching. In contrast, the $+c$ surface displays a significantly higher barrier, illustrating its strong resistance to alkaline etching. Additionally, we show that Ga–O–Ga bridges can form on etched surfaces, potentially serving as carrier traps. By providing a detailed atomistic understanding of GaN wet etching, this work offers valuable guidance for surface engineering in GaN-based device fabrication.

\end{abstract}


\twocolumn
\setlength{\parindent}{20pt}
\setlength{\columnsep}{0.7cm}
\justifying

\section{Introduction}

In recent decades, gallium nitride (GaN) has emerged as a versatile semiconductor, driving advancements in solid-state lighting technologies, such as light-emitting diodes (LEDs) and lasers, due to its exceptional optoelectronic properties as well as high chemical and thermal stability.\cite{Nakamura2009-vn,Simon2010-hp,Nakamura2013-vv,Choi2011-xq,Wasisto2019-lm,Iyer2020-aa} In addition, there is growing interest in leveraging GaN for next-generation power devices due to a high breakdown field of GaN, attributed to its wide band gap ($\sim$3.4 eV), and fast switching performance. Notably, GaN can generate a two-dimensional electron gas (2DEG) with high carrier mobility at its junction with AlGaN, enabling the development of high-electron-mobility transistors (HEMTs) for power applications.\cite{HEMT_vanDeurzen2024,HEMT_ieee_10098926,HEMT_wet_etching} 

Recently, the demand for miniaturizing GaN-based devices has been growing to support emerging display and electronic technologies.\cite{micro-LED_expt,green_micro-LED_Wu2024,science_GaN_microrod} For instance, virtual reality and augmented reality displays have become essential technologies in today's hyper-connected society.\cite{Park2019-fx,Park2021-oi} Achieving these displays requires extremely high pixel densities—exceeding 3,000 ppi—to eliminate screen-door effects,~\cite{Huang2020} necessitating the use of submicron-size LEDs based on precisely aligned GaN micro- or nanorods.\cite{Park2021-oi,Zhang2018-dk,Gou2019-bp} Moreover, GaN nanowires have the potential to extend the use of nanowire-based electronics beyond typical low-power applications—such as ultra-scaled digital circuits and 5G communications—to high-power applications like power conversion.\cite{Advanced_material_nanowire_review}  

The fabrication method of GaN nanostructures, particularly device-integrable one-dimensional forms such as nanowires and nanorods,\cite{Nanorod_D2NA00496H,Nanorod_RYU2024160040,Nanowire_Johnson2002}are classified into top-down and bottom-up approaches. Bottom-up techniques, such as molecular-beam epitaxy (MBE) and metal organic vapor phase epitaxy (MOVPE), are advantageous for achieving high crystal quality, including low dislocation densities and minimal lattice strain.\cite{Nanorod_RYU2024160040} However, nanostructures produced through bottom-up methods often exhibit undesired chemical and structural inhomogeneities, along with atomic-scale defects due to the use of molecular precursors.\cite{bottom-up_mocvd_doi:10.1021/acs.cgd.4c00937} Additionally, the consistent fabrication of uniformly aligned submicron-level geometries remains challenging using bottom-up approaches.\cite{top-down_Yulianto2021} 

Top-down approaches, which integrate lithography and etching processes, hold promise as an industrial method for mass production of wafer-scale GaN nanostructure arrays with precisely controlled shapes and dimensions.\cite{top-down_Oliva_2023} Typically, the etching process, which plays a key role in determining the shape of GaN nanostructures, consists of two steps: dry and wet etching. In dry etching, high-energy particles, such as \ce{Cl2}/Ar plasmas, directly bombard a pre-grown nitride film, tearing off atoms from the surface and enabling the rapid formation of GaN nanorods. However, this process can result in damaged, rough sidewalls with numerous defects, significantly degrading device performance.\cite{top-down_Oliva_2023,Step-flow_expt_JALOUSTRE2024108095}

Following dry etching, wet etching under alkaline environments, such as KOH and tetramethylammonium hydroxide solutions, is performed to refine the shape and size of nanostructures.\cite{Defect_SSPC_10.1063/5.0018829,wet_etching_tmah_treatment_APL,Step-flow_expt_JALOUSTRE2024108095} During wet etching, damaged surface layers are removed through chemical reactions between the nitride and etchant solution, promoting the formation of smooth, straight sidewalls. However, while wet etching mitigates surface damage caused by prior dry etching, it can also introduce other types of surface defects.\cite{Defect_SSPC_10.1063/5.0018829} Given the device performance based on extremely scaled GaN is highly sensitive to surface properties due to the high surface-to-volume ratio (e.g., rapid degradation of GaN micro-LEDs with decreasing LED size)\cite{EQE_OLIVIER2017112,EQE_Ley,ami_sidwall_defects}, a comprehensive understanding of GaN wet-etching processes is therefore essential for enhancing device performance.  

To date, various models have been proposed to elucidate the wet-etching behavior of GaN in alkaline solutions, with particular focus on the etching resistance of different surface orientations. For example, a previous study has suggested that etching resistance increases with the density of surface Ga and N ions, because the limited empty space on the surface hinders the attack of etchants, such as hydroxyl ions (\ce{OH-}), for chemical reactions.~\cite{CCA_GUO2023107173} Several groups have insisted that surfaces with a higher concentration of nitrogen ions possessing dangling bonds or lone-pair electrons are more resistant to alkaline etching because \ce{OH-} ions in solution experience greater electrostatic repulsion, making it more difficult for them to approach the surface.~\cite{review_ZHUANG20051,selective_etching_of_polar_surface,Tautz_etching_mechanism_oxidation,Tautz_pssa} Conversely, it has also been proposed that the presence of lone-pair electrons on surface nitrogen enhances wet etching by facilitating the subsequent adsorption of \ce{H+} onto nitrogen following Ga oxidation.~\cite{jpcc_etching_mechanism_lone_electron_pair} While these models may help explain the etching behavior of GaN surfaces under specific experimental conditions, they provide limited insight for surface engineering in GaN-based device fabrication due to the lack of detailed chemical reaction mechanisms. Additionally, beyond etching resistance, the evolution of surface morphology and the formation of surface defects during wet etching are critical issues for industrial applications. So far, these aspects have not been thoroughly investigated.

Ab initio molecular dynamics (AIMD) simulations based on density functional theory (DFT) have been widely employed to investigate surface chemical reactions at the atomic scale, owing to the high accuracy of DFT. However, this approach requires significant computational costs, limiting the simulation size and time. Recent advances of machine-learning potentials (MLPs), which are trained on DFT results, have offered promising alternatives to overcome the limitations of DFT.~\cite{MLP_Adv.Mat,MLIP_Fedik2022,MLP_Adv.Mat_2024_review} For example, molecular dynamics (MD) simulations using a Behler-Parrinello-type neutral network potential (NNP) have been employed to investigate ammonia decomposition on lithium imide surfaces, successfully explaining experimental observations and providing important insights into the catalytic reaction mechanism.~\cite{nature_catalyst_ammonia_decomposition} In addition, MLP-based MD simulations have been used to explore various chemical pathways for combustion of gases, gas-phase $\rm S_N2$ reaction, phosphoester bond formation and rupture in solution, and oxidization of Pt surface.~\cite{XING2024151492_combustion_NO2,MLIP_exploration_raction,PNAS_chemical_reactivity,JJS_doi:10.1021/acs.jctc.4c00767}

 
In this work, we perform large-scale NNP-MD simulations to investigate wet etching of GaN surfaces, including two polar surfaces ($+c$ and $-c$) and two non-polar surfaces ($a$ and $m$), in KOH solution, which are important for various industrial applications. By training on comprehensive DFT datasets and iteratively updating model parameters, we develop a Behler-Parrinello-type NNP capable of accurately describing chemical reactions between GaN and KOH solution across a wide range of temperature and pressure conditions. To simulate the wet etching of GaN, we perform NNP-MD simulations using the temperature-accelerated dynamics (TAD) approach under elevated temperature and pressure conditions, which accurately reproduces the experimentally observed structural modification of a GaN nanorod during alkaline etching. We examine the evolution of surface morphology during wet etching, which reveals that pyramidal etch pits form on the $-c$ surface, while truncated pyramidal pits develop on the $+c$ surface, exposing facets such as $\{10\bar{1}\bar{1}\}$ planes. On the non-polar surfaces, etch pits grow laterally, resulting in planar etched morphologies that retain the original surface orientation. From the analysis of MD trajectories, we identify key chemical reactions that constitute the etching mechanisms of GaN surfaces. We perform enhanced-sampling MD simulations for Ga dissolution on each surface, a critical step in determining the etching rate, constructing the corresponding free-energy profiles under realistic etching conditions. The results show that the $-c$, $a$, and $m$ planes exhibit moderate activation energies, highlighting high feasibility of wet etching. In contrast, the $+c$ plane yields a prohibitively high energy barrier, indicating the difficulty of its alkaline etching. We also demonstrate that \ce{Ga-O-Ga} bridges, which would serve as surface defects detrimental to device performance, can form on etched surfaces of GaN. 

\section{Results and discussion}
\subsection{Training neural network potential}

\begin{figure*}
    \centering
    \includegraphics[width=17.8cm]{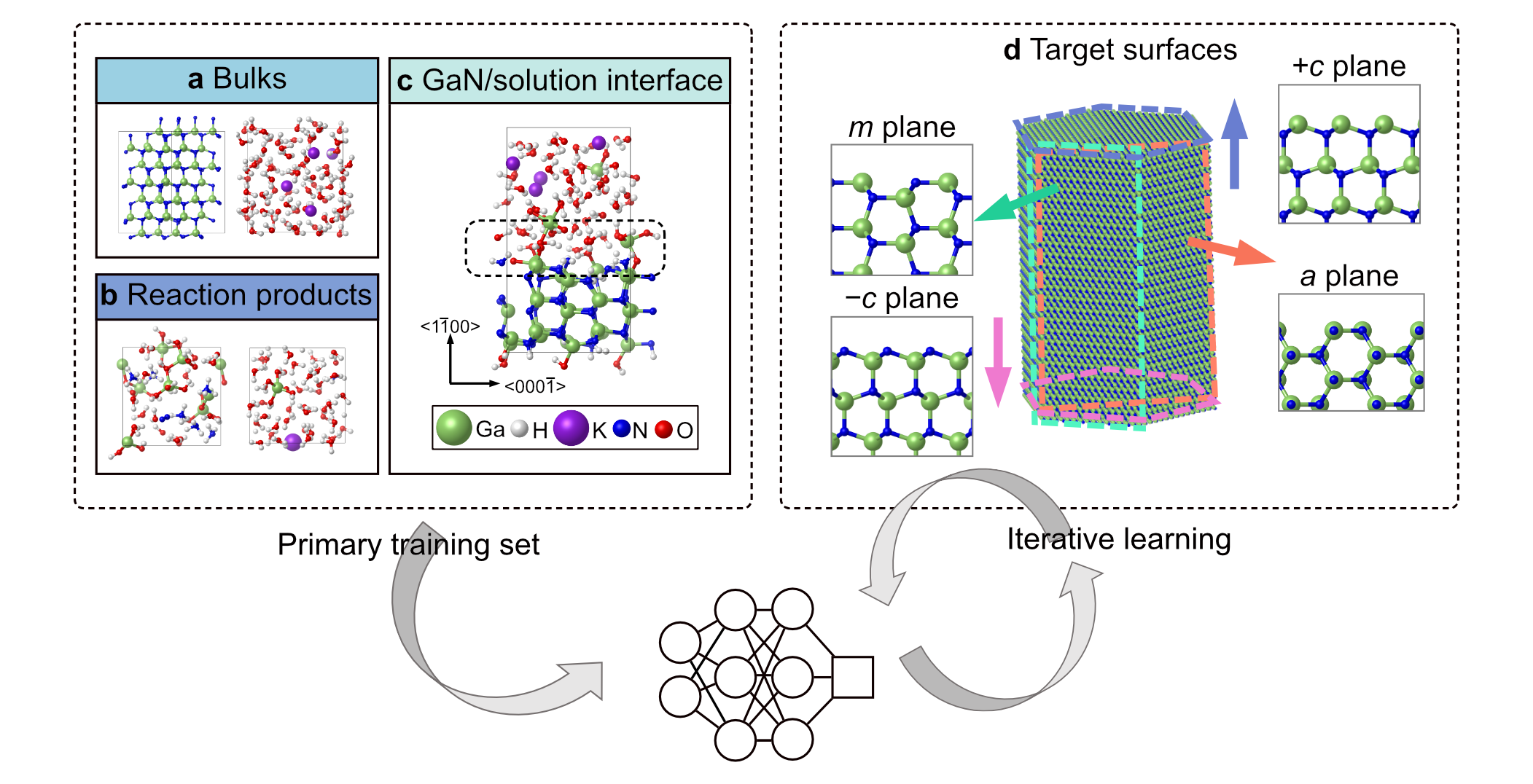}
    \caption{Schematic illustration of the training dataset construction and iterative learning procedure adopted in this study. (a-c) Components comprising the primary training set for the baseline NNP model. (d) Target surfaces considered during iterative learning.}
    \label{fig:Training set and iterative learning for NNp}
\end{figure*}

To develop a NNP capable of simulating the wet etching of various GaN crystal surfaces, including the $+c$, $-c$, $a$, and $m$ planes, it is essential to construct a comprehensive training set that encompasses not only the bulk properties of GaN and alkaline solutions but also a wide range of relevant chemical reactions at their interface. However, due to the immense computational cost, it is infeasible to identify all possible chemical reactions exhaustively. To address this challenge, we first construct a baseline NNP model from a primary training set and subsequently refine it through iterative updates of model parameters, as illustrated in Figure~\ref{fig:Training set and iterative learning for NNp}. 

\subsubsection{Primary training dataset}{\noindent}
Table~\ref{tab:configurations} provides an overview of the primary training dataset, which is categorized into three components: bulk structures, reaction products, and GaN/solution interfaces. 

\begin{table*}
\renewcommand{\arraystretch}{1.5}
    \caption{Configurations included in the complete training dataset, encompassing both the primary training set and additional structures incorporated during iterative learning}
    \centering
    \resizebox{\textwidth}{!}{\begin{tabular}{c|c|c|c|c}
    \hline
    Category & \multicolumn{2}{c|}{Structure type} & Number of structure & Number of training points \\ \hline
    \multirow{2}{*}{Bulks} 
        & \multicolumn{2}{c|}{GaN} &  474  & 6,884  \\
        & \multicolumn{2}{c|}{KOH} &  1,222 & 224,874  \\ \hline
    \multirow{7}{*}{Reaction products} 
        & \multicolumn{2}{c|}{\ce{Ga2O3}+\ce{NH3}} & 414 & 53,820  \\
        & \multicolumn{2}{c|}{\ce{Ga(OH)3}+\ce{NH3}} & 414 & 45,540   \\
        & \multicolumn{2}{c|}{\ce{Ga(OH)3}+\ce{NH3}+$m$\ce{H2O}} & 414 & 46,368   \\
        & \multicolumn{2}{c|}{\ce{[Ga(OH)4]-}+\ce{K+}+\ce{NH3}} & 414 & 74,934 \\
        & \multicolumn{2}{c|}{\ce{[Ga(OH)4]-}+\ce{K+}+\ce{NH3}+$m$\ce{H2O}} & 207 & 27,738 \\
        & \multicolumn{2}{c|}{\ce{[Ga(OH)4]-}+\ce{K+}+\ce{H2O}} & 414 & 55,890 \\
        & \multicolumn{2}{c|}{\ce{NH3}+\ce{K+}\ce{OH-}+\ce{H2O}} & 414 & 53,406  \\ \hline   
    \multirow{3}{*}{GaN/solution interface} 
        & \multirow{2}{*}{$m$-plane} & Top & 936 & 156,312  \\
        &   & Sub & 936  & 156,312  \\ \cline{2-5}
        & \multicolumn{2}{c|}{Iterative learning} & 5,308 & 1,554,126 \\ \hline
    \multicolumn{3}{c|}{Total} & 11,304 & 2,463,556 \\ \hline
    \end{tabular}}
    \label{tab:configurations}
\end{table*}

(1) Bulks: For GaN, the training data include configurations from wurtzite crystal, amorphous, and liquid phases, as illustrated in Figure~\ref{fig:Training set and iterative learning for NNp}a. To capture diverse local geometries, MD simulations are conducted across various temperatures. Specifically, for the crystal phase, we prepare a 3$\times$3$\times$3 supercell of a perfect wurtzite GaN crystal, along with supercells containing either Ga or N vacancies. Trajectories for each supercell are sampled from NVT MD simulations performed at 1500 K for 3 ps. For the liquid phase, 40 Ga and 40 N atoms are initially distributed randomly within a supercell with the volume corresponding to the experimental crystal density of GaN (6.15 g/cm$^3$). The structure is first premelted using NVT MD simulations at 4000 K, above the melting point of GaN, without sampling. Subsequently, the liquid is simulated at 3000 K for 10 ps, during which configurations are sampled for the training set. To obtain amorphous configurations, the liquid structure is quenched to 300 K at a rate of 150 K/ps and then annealed for 4 ps at 300 K. Configurations are sampled during the quenching and annealing steps. 

For KOH solutions, we generate a supercell containing 4 KOH and 50 H$_2$O molecules, corresponding to a 4 M molar concentration under ambient conditions (1 bar and 350 K). The in-plane lattice parameters are fixed to match a 3$\times$2 extension of the GaN $m$-plane lattice, enabling seamless integration with subsequent GaN/solution interface simulations, while the $z$ component is allowed to relax. We conduct NPT MD simulations for 8 ps across a broad range of pressures (1 bar and 100 kbar) and temperatures (350, 600, and 2000 K). The high-pressure (100 kbar) and high-temperature (2000 K) conditions are considered to reflect those used in etching simulations based on the TAD approach (see the Methods section for details on the etching simulation). Snapshots from these simulations are sampled to construct the training set. On the other hand, due to the finite size of the supercell, its volume fluctuates to some extent during NPT simulations, leading to variations in solute concentration between 3 M and 7 M. To sample molecular configurations at a consistent concentration, we also perform NVT simulations, setting the $z$ lattice parameter to the time-averaged value obtained from the NPT simulation. We confirm that the time-averaged $z$ lattice at 1 bar and 350 K leads to a solution concentration of 4 M, aligning with experimental results. It is worth noting that, although the high pressure and temperature conditions for etching simulations drive the solution into a supercritical state, the density of the solution (1.57 g/cm$^3$) is close to that at 1 bar and 350 K (1.18 g/cm$^3$). Furthermore, while the self-ionization of water is slightly enhanced under these conditions compared to ambient conditions, the constituents (H$_2$O, K$^+$, and OH$^-$) remain stable throughout the MD simulations, without formation of radical species.

(2) Reaction products: To sample aqueous molecular configurations of potential byproducts formed during alkaline wet etching, we perform NVT MD simulations. The types and amounts of elements initially included in each supercell are determined considering five chemical reactions between GaN and aqueous solutions with or without KOH, targeting specific byproducts (see detailed procedure in Section S2.1 in the Supporting Information). Moreover, we additionally sample configurations of \ce{[Ga(OH)4]-} and \ce{NH3} in KOH solution, because these species are thermodynamically more favorable and are therefore expected to form more readily during wet etching compared to other byproducts.  

(3) GaN/solution interface: To generate atomic configurations at the GaN/solution interface for the primary training set, we focus on the $m$ plane, as its surface structure is more complex than those of the other planes (such as $a$ and $\pm c$ surfaces). This structural complexity allows for the generation of diverse local environments during MD simulations, potentially covering the structural characteristics of the other planes to some extent. The $m$ plane exhibits a bilayer structure: Ga and N ions in the top layer possess a single dangling bond, whereas those in the sublayer have two dangling bonds upon cleavage (Figure~\ref{fig:Training set and iterative learning for NNp}d). The interface model consists of a GaN slab, with either the top layer or sublayer exposed, in contact with an alkaline solution containing 4 KOH and 50 H$_2$O molecules. Prior to the MD simulations, Ga and N dangling bonds exposed to the solution are passivated with OH and H species, respectively—a process that occurs spontaneously in water due to the energetic instability of the dangling bonds.~\cite{acs_catlyst_water_dissociation_m-plane}
We sample trajectories associated with Ga and N dissolution from NVT MD simulations performed for 5 ps at 100 kbar and elevated temperatures (2700 K and 3000 K for the top-layer model and 2000 K for the sublayer model). The use of the lower temperature for the sublayer model reflects its higher reactivity, and therefore, a faster etching rate. During these simulations, the in-plane lattice parameters of the interface models are set to those of the GaN slab, considering the rigidity of the GaN lattice. The $z$ component is determined as the time-averaged value obtained from the last 3 ps of the preceding NPT MD simulation (see Section S2.2, Supporting Information, for details).  

\subsubsection{Iterative learning} \label{sec:iterative_learning}
\begin{figure}
    \centering
    \includegraphics[width=8.5cm]{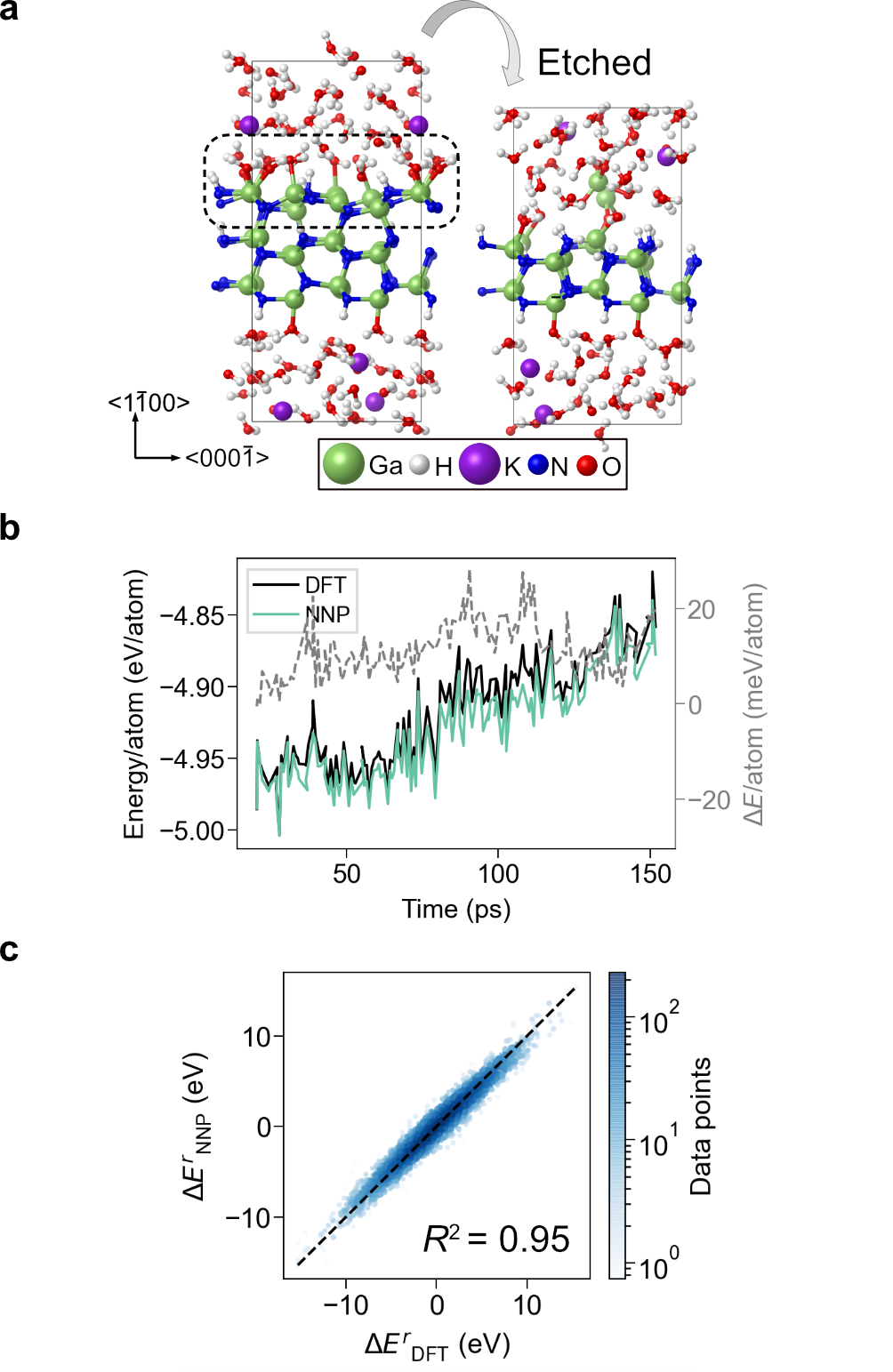}
    \caption{Validation of the refined neural network potential (NNP) after three iterations of learning. (a) Structural snapshots showing the progression of wet etching on the GaN $m$ plane, from the pristine surface to a state with one bilayer removed, obtained from a 150 ps molecular dynamics simulation at 2000 K and 100 kbar. (b) Temporal evolution of atomic energies calculated by the NNP and DFT. (c) Parity plot of reaction energies comparing NNP predictions with DFT calculations. The snapshots used in (b) and (c) were selected from the etching trajectory using a graph-based filtering method.}
    \label{fig:Iterative learning results:E and F}
\end{figure}

We first validate the accuracy of the baseline NNP, trained on the primary dataset, by comparing its predictions with DFT results for interface properties and etching behavior of the GaN surfaces. To this end, we perform NNP-MD etching simulations on polar ($+c$ and $-c$ plane) and non-polar ($m$ and $a$ plane) surfaces within the TAD approach at 2000 K and 100 kbar. To ensure the feasibility of DFT calculations, the simulation cell size is restricted to include approximately 300 atoms. The supercells are constructed such that only the upper surface of the GaN slab interacts with the solution, while the bottom surface remains unreactive (details on the passivation of the bottom surface are provided in Section S3.1, Supporting Information). 
Etching simulations are terminated once a single GaN layer (a bilayer in the case of $m$ surface) completely dissolves, as shown in Figure \ref{fig:Iterative learning results:E and F}a for the $m$ surface. 

Subsequently, we evaluate DFT energies for selected snapshots to estimate the prediction errors of the baseline NNP. Since etching involves chemical reactions that reorganize chemical bonds, it is crucial for NNP to accurately describe the reaction energy, i.e, the energy difference before and after chemical-bond reorganization. Hence, we identify reaction moments by analyzing the MD trajectories using a graph-based analysis (Figure S3) and select snapshots before and after the chemical reactions for error estimation. We observe that the baseline NNP produces energy errors exceeding 30 meV/atom for some snapshots, indicating that local configurations encountered during etching simulations are occasionally not well represented by the primary training set. The worst case appears for the $-c$ plane, where the maximum error exceeds 100 meV/atom (Figure S4). To address these large errors (greater than 30 meV/atom), we add the corresponding snapshots to the training set and retrain the model. After three iterations of this refinement process, our NNP model achieves energy errors below 30 meV/atom for all surface orientations, including the $-c$ plane, as shown in Figure~\ref{fig:Iterative learning results:E and F}b. In particular, it is worth noting that the iteratively trained NNP accurately reproduces DFT reaction energies, yielding a coefficient of determination ($R^2$) of 0.95 for their correlation (Figure~\ref{fig:Iterative learning results:E and F}c). This underlines the reliability of the model for etching simulations. In addition to the reduction of energy errors, the iterative learning process significantly decreases force errors, thereby enabling stable MD simulations of the etching process (Figure S5).

The precision of the iteratively refined NNP is thoroughly validated for various fundamental properties of bulk GaN and KOH solutions, such as bulk moduli, equation of state, and diffusivity (see Sections S4.1 and S4.2 in Supporting Information for details). Additionally, we confirm that the NNP accurately captures molecular arrangements at the GaN/solution interface. For example, MD simulations at 1 bar and 300 K show the accumulation of H$_2$O molecules near the interface, with their O ions oriented toward H ions on the GaN surface, forming hydrogen bonds (Figure S8). Furthermore, both the density and orientational distribution of water molecules gradually recover those of the bulk with increasing distance from the interface.  

\subsection{Structural modification of GaN nanorod by alkaline etching}

\begin{figure*}
    \centering
    \includegraphics[width=17.8cm]{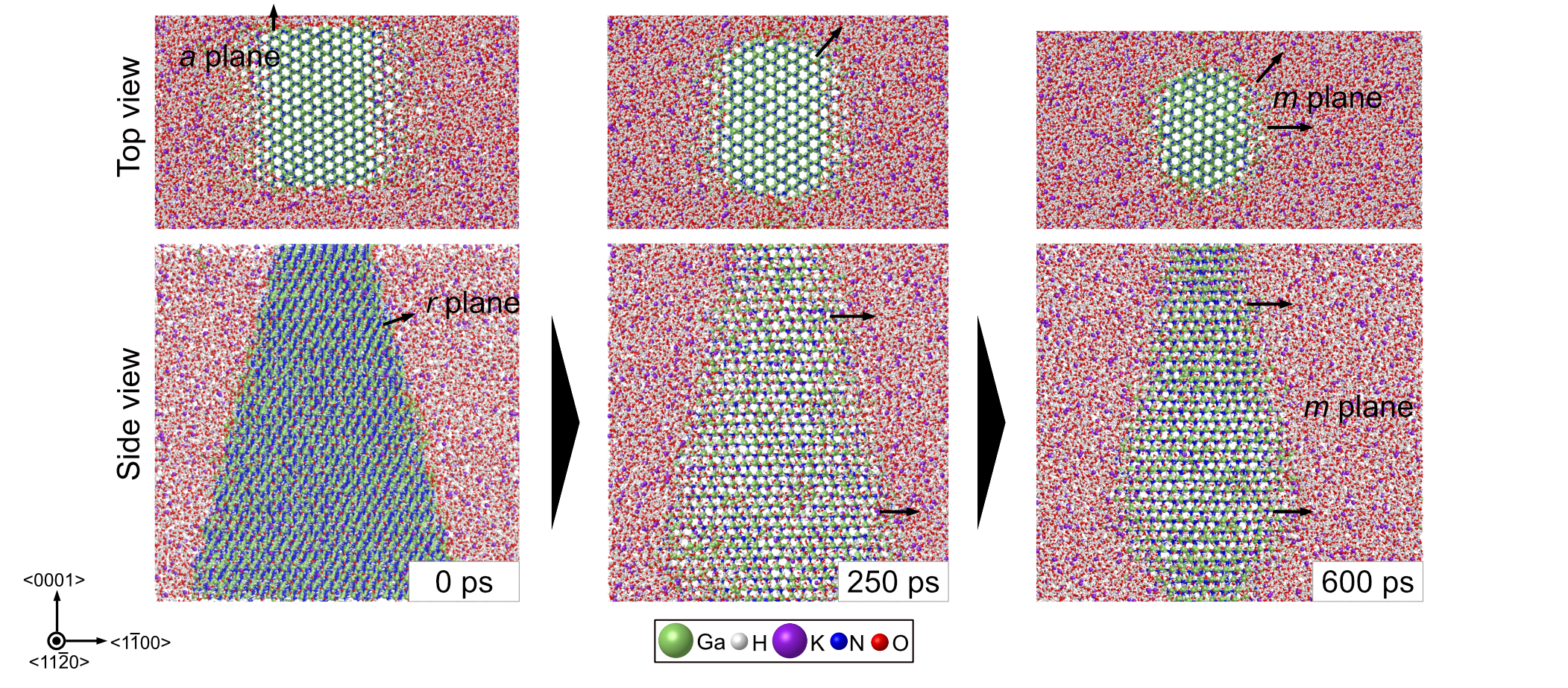}
    \caption{Simulated structural evolution of a GaN nanorod during alkaline wet etching at 2000 K and 100 kbar.}
    \label{fig:tree}
\end{figure*}

To demonstrate the accuracy of etching simulations based on the refined NNP, we examine the structural evolution of a GaN nanorod under alkaline etching. The NPT MD simulation is carried out for 600 ps using the TAD approach at 2000 K and 100 kbar, with the solution pH set to 14. The height of the nanorod is maintained by fixing the $z$-axis during the simulation. The GaN nanorod model, consisting of approximately 50,000 atoms, initially adopts a truncated-pyramid structure (Figure~\ref{fig:tree}), exposing the $a$ and $r$ planes on its sidewalls. As etching proceeds, the slope of the original rod gradually disappears, and the structure evolves into a hexagonal shape with relatively flat sidewalls dominated by the $m$ plane, which exhibits a slower etching rate compared to the $a$ and $r$ planes. The simulation results align well with previous experiments, in which slanted nanorods formed by dry etching gradually transform into a hexagonal shape, exposing the $m$ plane during alkaline wet etching.~\cite{Nanorod_D2NA00496H,Nanorod_RYU2024160040,top-down_Yulianto2021,nanorod_truncated-pyramid_CHEN2015168} This underscores the capability of the accelerated NNP-MD approach adopted in this study to accurately simulate the wet etching process. 

\subsection{Morphologies of etched surfaces}

In this section, we discuss the morphologies of etched surfaces that are generated from MD simulations leveraging the refined NNP within the TAD approach. The GaN/solution interface models contain thousands of atoms, including about 1,000 GaN formula units. The solution models are constructed to exhibit a pH of approximately 14. The specific information about the supercell models are shown in Table S3. Prior to MD simulations, Ga and N dangling bonds of pristine surface models are passivated with OH and H species, respectively. All models are initially equilibrated for 100 ps at 20 kbar and 1000 K. Etching simulations are then carried out at 100 kbar and 2000 K for several hundred picoseconds.

\subsubsection{Polar surfaces}

\begin{figure*}
    \centering
    \includegraphics[width=17.8cm]{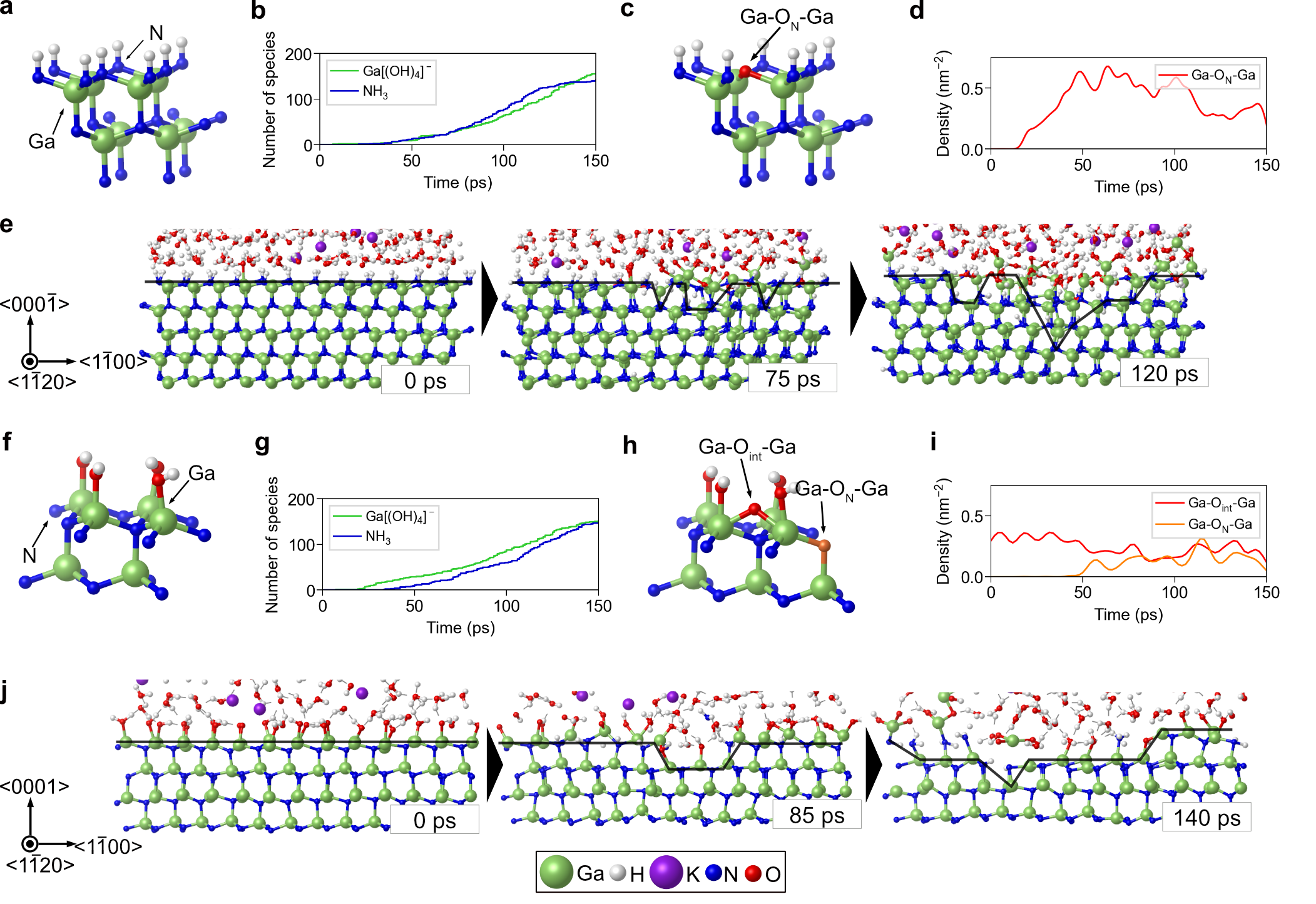}
    \caption{Results of NNP-MD etching simulations for polar GaN surfaces: (a-e) the $-c$ plane and (f-j) the $+c$ plane. (a, f) Atomic structures of the pristine $-c$ and $+c$ surfaces. (b, g) Cumulative number of dissolved species over time for the $-c$ and $+c$ surfaces. (c, d; h, i) Representative oxygen bridge structures and their areal densities at time $t$ for the $-c$ and $+c$ surfaces, respectively. (e, j) Temporal evolution of etched morphologies for the $-c$ and $+c$ surfaces.}
    \label{fig:polar}
\end{figure*}

We first discuss the $-c$ plane, for which surface morphologies formed by wet etching have been extensively studied in experiments.\cite{EBI_Lai,Fujii_LED_N-face} The slab model for the $-c$ plane prior to etching is shown in Figure~\ref{fig:polar}a. Figure~\ref{fig:polar}b depicts the cumulative number of the major etching products, \ce{[Ga(OH)4]-} and \ce{NH3}, as a function of time. At the beginning of the simulation, Ga and N ions on the surface are progressively decorated with OH$^-$ and H$^+$, respectively, with negligible formation of the etching products. Once Ga and N dissolution begins at certain surface sites, nearby surface ions become destabilized due to the loss of \ce{Ga-N} bonds, thereby accelerating the etching process. As a note, in the alkaline solution, the main source of H$^+$ ions is H$_2$O, which can dissociate into H$^+$ and OH$^-$ at the interface. Upon H$_2$O dissociation, H$^+$ adsorbs onto a surface N ion, while the OH$^-$ ion mostly diffuse into the solution. Interestingly, N dissolution proceeds at a rate comparable to that of Ga dissolution, even in the alkaline solution where protons are scarce. This behavior is attributed to the higher positions of N ions relative to Ga ions, which enhances the accessibility of \ce{H2O} to N ions, partially compensating for the limited availability of H$^+$ ions in the alkaline environment. 

While the etching proceeds, we occasionally observe the formation of \ce{Ga-O-Ga} bridges (Figure~\ref{fig:polar}c) as intermediates, as shown in Figure~\ref{fig:polar}d. These \ce{Ga-O-Ga} units are formed when an OH$^-$ ion replaces a N ion during N dissolution. This is followed by the removal of an H$^+$ ion through a reaction with a neighboring OH$^-$ ion, resulting in the formation of H$_2$O. Note that the formation of oxygen bridges is not a prerequisite for N dissolution, as inferred by their low areal density during the etching (Figure~\ref{fig:polar}d). However, once formed, they can persist for a long period. The thermodynamic stability of oxygen bridges and their influence on the etching rate will be discussed in detail later. 

The temporal evolution of the surface morphology of the $-c$ plane is illustrated in Figure~\ref{fig:polar}e. Initially, pyramidal etch pits form at several locations on the surface. These pits then undergo lateral expansion along specific directions, such as <11$\bar{2}$0> and <1$\bar{1}$00>, leading to their widening. Following a short period of lateral etching that exposes atomic configurations resembling those of the initial surface, vertical etching proceeds concurrently, resulting in the formation of deeper pits. Consequently, the etch pits grow three-dimensionally over time, while retaining their characteristic pyramidal shape. Considering the directions of the lateral and vertical etching, the exposed surfaces in etch pits are close to the $\{10\bar{1}\bar{1}\}$ plane. This observation aligns with experimental studies.\cite{review_ZHUANG20051,Tautz_etching_mechanism_oxidation,Tautz_pssa} 

Figure~\ref{fig:polar}f shows the surface model of the $+c$ plane used for the etching simulation. Unlike the $-c$ plane, preferential dissolution of Ga ions is pronounced (Figure~\ref{fig:polar}g), because N ions are positioned below Ga ions. As etching progresses, Ga–O–Ga bridges are formed (Figures~\ref{fig:polar}h and \ref{fig:polar}i). Two distinct configurations of these bridges are observed: (1) a \ce{Ga-O$_{\rm{N}}$-Ga} bridge, in which an O ion occupies a N site, linking the upper and lower Ga ions, and (2) a \ce{Ga-O$_{\rm{int}}$-Ga} bridge, where an O ion resides at an interstitial site between two surface Ga ions. The latter configuration, which does not require N dissolution, can form even during the pre-equilibration step, leading to a finite number of \ce{Ga-O-Ga} bonds at $t=0$ (Figure~\ref{fig:polar}i). Similar to the $-c$ plane, etch pits on the $+c$ plane initially expand laterally. Although etching in the vertical <000$\bar{1}$> direction is observed, it is less pronounced than that of the $-c$ plane (Figure~\ref{fig:polar}j). As a result, the surface morphology after wet etching is expected to resemble a truncated pyramid. To the best of our knowledge, there are no experimental reports on the morphologies of etched $+c$ surfaces in alkaline solutions, likely due to the difficulty of etching at typical process temperatures of 50–90$^\circ$C.\cite{review_ZHUANG20051,selective_etching_of_polar_surface} Further experimental studies at elevated temperatures are needed to validate our prediction on the morphology of etched $+c$ surfaces.   

\subsubsection{Non-polar surfaces}

\begin{figure*}
    \centering
    \includegraphics[width=17.8cm]{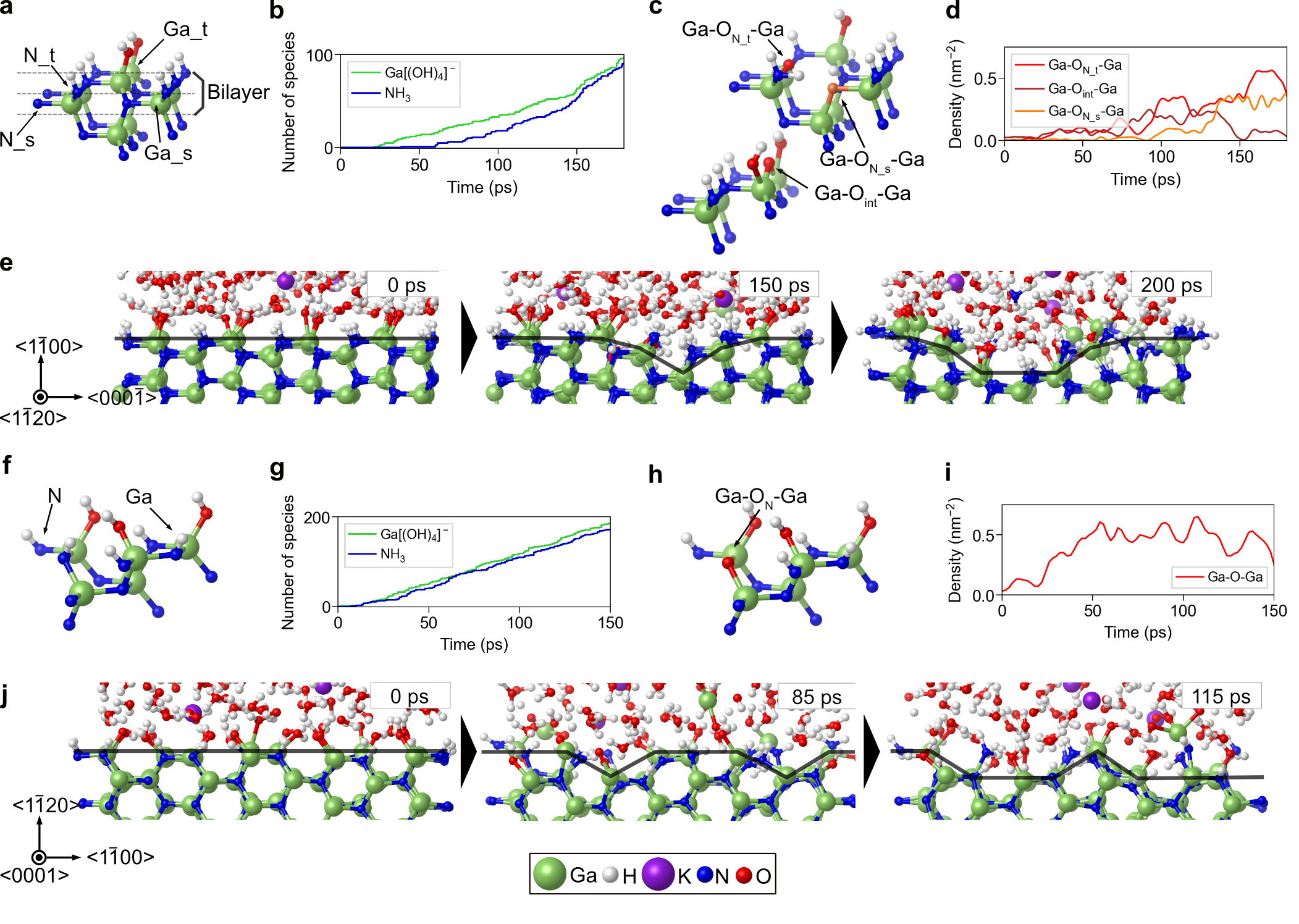}
   \caption{Results of NNP-MD etching simulations for non-polar GaN surfaces: (a-e) the $m$ plane and (f-j) the $a$ plane. (a, f) Atomic structures of the pristine $m$ and $a$ surfaces. (b, g) Cumulative number of dissolved species over time for the $m$ and $a$ surfaces. (c, d; h, i) Representative oxygen bridge structures and their areal densities at time $t$ for the $m$ and $a$ surfaces, respectively. (e, j) Temporal evolution of etched morphologies for the $m$ and $a$ surfaces}
    \label{fig:nonpolar}
\end{figure*}

The etching simulation for the $m$ plane begins with the atomic model illustrated in Figure~\ref{fig:nonpolar}a. It is clearly seen in Figure~\ref{fig:nonpolar}b that Ga dissolution occurs first, followed by N dissolution, despite surface Ga and N ions being located at the same height. This behavior can be attributed to the abundance of OH$^-$ ions in the alkaline solution, which promotes the formation of $[\rm{Ga(OH)_4}]^-$. Furthermore, N dissolution on the $m$ plane would be further delayed because each surface N ion has only a single nearest Ga neighbor in the top layer. This limited coordination with Ga is ineffective for top-layer N ions to increase \ce{N-H} bonds upon the dissolution of top-layer Ga ions, thereby slowing down the formation of \ce{NH3}. In the following section on mechanistic analysis, we will provide a more detailed discussion of the \ce{N-H} bond formation process.  

During the wet etching process, \ce{Ga-O-Ga} bridges are observed, as seen on the polar surfaces. Two \ce{Ga-O$_{\rm{N}}$-Ga} and one \ce{Ga-O$_{\rm{int}}$-Ga} configurations are identified, as depicted in Figure~\ref{fig:nonpolar}c. Between the \ce{Ga-O$_{\rm{N}}$-Ga} configurations, one involves an O ion replacing a N ion at the top surface (N$\_$t), and thus, the oxygen connects two sublayer Ga ions (Ga$\_$s). In the other configuration, an oxygen ion occupies a N site in the sublayer (N$\_$s), forming a bridge between a top-layer Ga (Ga$\_$t) ion and a sublayer Ga ion (Ga$\_$s). On the other hand, oxygen can occupy an atomic site between two top-layer Ga ions (Ga$\_$t), producing a \ce{Ga-O$_{\rm{int}}$-Ga} configuration. A finite number of \ce{Ga-O$_{\rm{int}}$-Ga} configurations, which do not involve N dissolution processes, keeps observed throughout the etching simulation even at $t=0$. In contrast, Ga-O$_{\rm{N}}$-Ga configurations emerge after N dissolution takes place.

Figure~\ref{fig:nonpolar}e depicts the morphological evolution of the $m$ plane during the wet etching. Etch pits initially form at several locations on the surface due to the dissolution of Ga$\_$t and N$\_$t ions. These pits preferentially grow linearly along the <11$\bar{2}$0> direction. Over time, the pits extend further along the <0001> direction as Ga$\_$s and N$\_$s ions dissolve in addition to Ga$\_$t and N$\_$t ions. Notably, etching in the downward direction does not happen during the lateral growth of the etch pits, highlighting the highly anisotropic nature of wet etching of the $m$ plane. Consequently, the etched surface adopts a planar morphology while retaining the surface orientation of the $m$ plane. These simulation results are consistent with experimental observations, which reported that wet etching of the $m$ plane produces a planar surface morphology rather than a pyramidal one, preserving the original surface orientation.\cite{HEMT_wet_etching,Step-flow_expt_JALOUSTRE2024108095}

The initial atomic structure of the $a$ plane before wet etching is shown in Figure~\ref{fig:nonpolar}f. Similar to the $m$ plane, Ga ions on the $a$ plane dissolve first, followed by dissolution of N ions (Figure~\ref{fig:nonpolar}g). However, N dissolution on the $a$ plane occurs more rapidly than that on the $m$ plane. This behavior can be attributed to the fact that N ions on the $a$ plane are bonded to two surface Ga ions, unlike those on the $m$ plane. As a result, it is more feasible for surface N ions to increase \ce{N-H} bonds following the dissolution of surface Ga ions. Intermediate \ce{Ga-O-Ga} bridges are observed during the etching of the $a$ plane, as shown in Figures~\ref{fig:nonpolar}h and i. These bridges adopt a \ce{Ga-O$_{\rm{N}}$-Ga} configuration, where an oxygen ion replaces a nitrogen site, bridging a Ga ion exposed to the solution with an underlying Ga ion. 

The morphology and growth patterns of the etched surface of the $a$ plane are similar to those of the $m$ plane. During the initial stages of etching, etch pits are generated on the surface through the dissolution of Ga and N ions exposed to the solution, as shown in Figure~\ref{fig:nonpolar}j. These pits first extend along the <0001> direction. Subsequently, they grow along the <1$\bar{1}$00> direction without notable vertical growth. As a result, the etched surfaces retain planar morphologies, preserving the original surface orientation. These findings align with previous experimental reports.\cite{HEMT_wet_etching,Step-flow_expt_JALOUSTRE2024108095} 

\begin{figure}
    \centering
    \includegraphics[width=8.5cm]{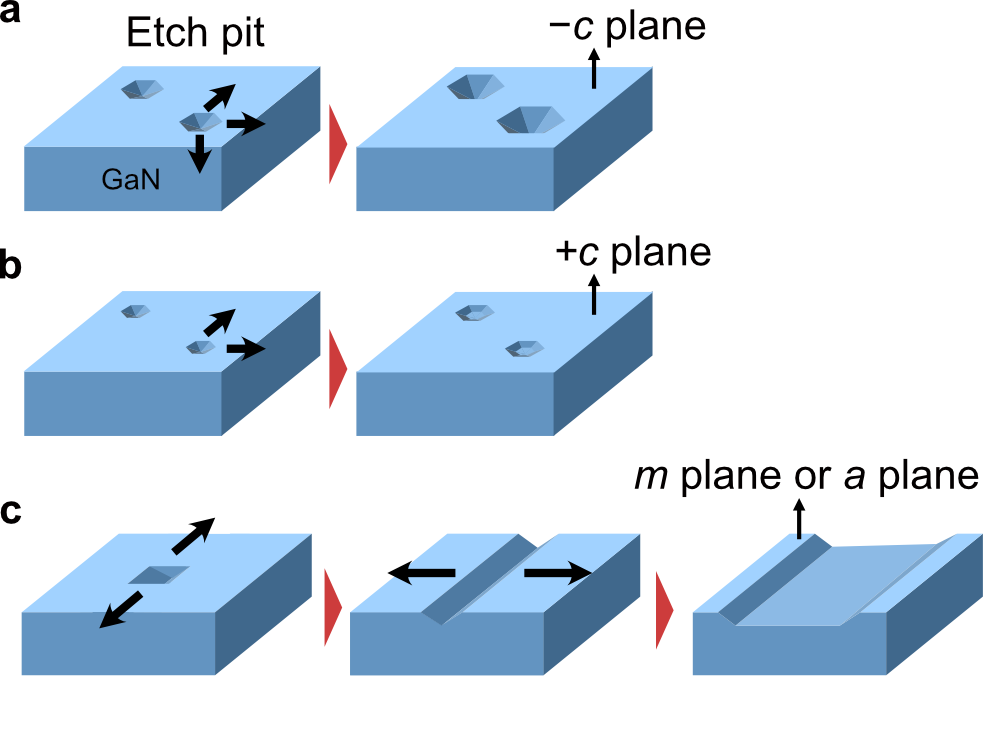}
    \caption{Schematic illustration of etch-pit growth on (a) the $-c$ plane, (b) the $+c$ plane, and (c) nonpolar surfaces ($m$ and $a$ planes).}
    \label{fig:etching_schematic}
\end{figure}

Based on the foregoing discussions, the temporal evolution of etched surfaces on both polar and non-polar GaN planes is illustrated in Figure~\ref{fig:etching_schematic} .  

\subsection{Etching mechanism}

\begin{figure*}
    \centering
    \includegraphics[width=17.8cm]{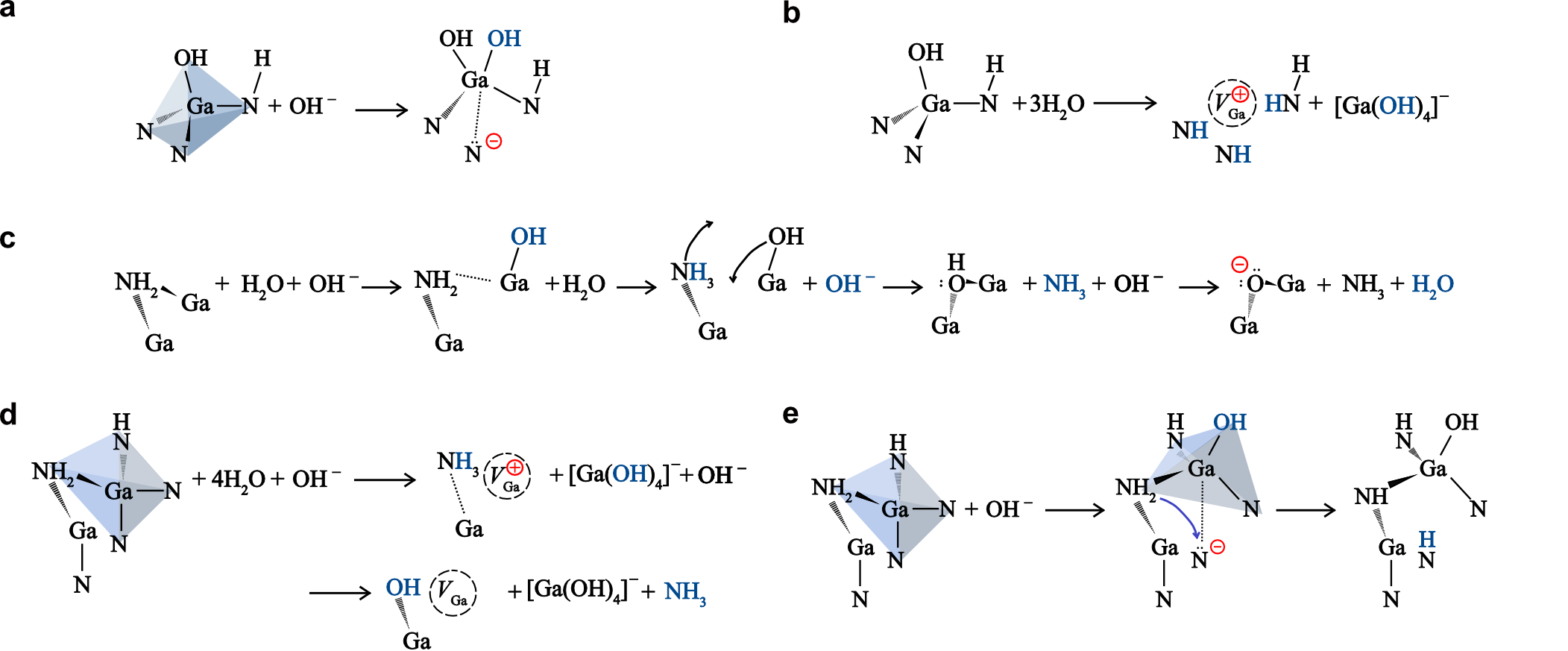}
    \caption{Key etching processes identified from NNP-MD simulations: (a) Adsorption of \ce{OH-} leading to \ce{Ga-N} bond breaking, (b) Ga dissolution, (c) N dissolution with and (d) without the formation of \ce{Ga-O-Ga} bridges, and (e) \ce{Ga-N} bond breaking facilitated by proton transfer from \ce{-NH2} species. In (a)-(e), reaction products and adsorbates are highlighted in blue. $V_{\rm{Ga}}$ denotes a Ga vacancy.}
    \label{fig:mechanism_pathways}
\end{figure*}

Through detailed analysis of the MD trajectories, we identify key etching processes that commonly occur across all GaN surface orientations (Figure~\ref{fig:mechanism_pathways}). First, the \ce{OH-} adsorption onto a Ga ion, which initiates the etching process, causes a significant upward displacement of the Ga ion due to the attractive interaction between the attached \ce{OH-} ions and the Ga ion (Figure~\ref{fig:mechanism_pathways}a). This structural distortion progressively weakens the \ce{Ga-N} bonds on the surface as the number of attached \ce{OH-} ions increases. Consequently, one of the \ce{Ga-N} bonds breaks, leading to an electron lone-pair on the N ion. In subsequent reactions, this lone-pair state is passivated by a \ce{H+} ion. 

Second, as shown in Figure~\ref{fig:mechanism_pathways}b, Ga dissolution, which requires the consecutive breaking of \ce{Ga-N} bonds, leaves behind \ce{-NH} or \ce{-NH2} species near Ga vacancy sites. The \ce{-NH2} species can be converted into \ce{NH3} in the next step, leading to N dissolution. There are two pathways for N dissolution. The first one involves the formation of a \ce{Ga-O-Ga} bridge (Figure~\ref{fig:mechanism_pathways}c). Specifically, the attack of \ce{OH-} on a Ga ion bonded to the \ce{-NH2} species results in the formation of a \ce{Ga-OH} bond. This subsequently breaks the \ce{Ga-N} bond, leading to the immediate protonation of the \ce{-NH2} via the dissociation of an incoming \ce{H2O} molecule. Once the \ce{NH3} molecule dissolves, the remaining nitrogen vacancy is occupied by the attached \ce{OH-} ion, forming a bridge with a nearby Ga ion. Afterward, the hydrogen ion of the bridging \ce{OH-} is dissociated through a reaction with another \ce{OH-} ion in the solution, producing \ce{H2O}. On the other hand, if breaking the Ga\ce{-NH2} bond is energetically unfavorable, further Ga dissolution occurs first, and \ce{NH3} spontaneously dissolves, as shown in Figure~\ref{fig:mechanism_pathways}d. Between these two pathways for N dissolution, the latter, which proceeds without the formation of an oxygen bridge, occurs more frequently. Nonetheless, a moderate amount of oxygen bridges is expected to be present on the etched surface due to their kinetic stability, which will be discussed later.

Third, \ce{-NH2} species can serve as a proton carrier, facilitating proton transfer to a underlying N ion that is otherwise difficult to gain a proton directly from a H$_2$O molecule (Figure~\ref{fig:mechanism_pathways}e). Once this transfer occurs, the reverse reaction, namely the separation of Ga and \ce{OH-}, becomes energetically unfavorable. As a result, the \ce{Ga-OH} bonds can be sustained for a long duration, thereby promoting the dissolution of Ga ions. 

In the following, we present free-energy profiles for Ga dissolution along with associated structural changes and chemical reactions on each surface. As discussed in Section 2.2, the onset of GaN etching is delayed until the dissolution of several Ga ions occurs. This underscores the critical role of Ga removal from the pristine surface, a process that recurs throughout wet etching, in determining the overall etching rate. The free-energy profiles are obtained via $On-the-fly$ $probability$ $enhanced$ $sampling$ (OPES) simulations at 1 bar and 350 K, corresponding to experimental pressure and temperature conditions. 

\subsubsection{Ga dissolution on polar surfaces}
\begin{figure*}
    \centering
    \includegraphics[width=17.8cm]{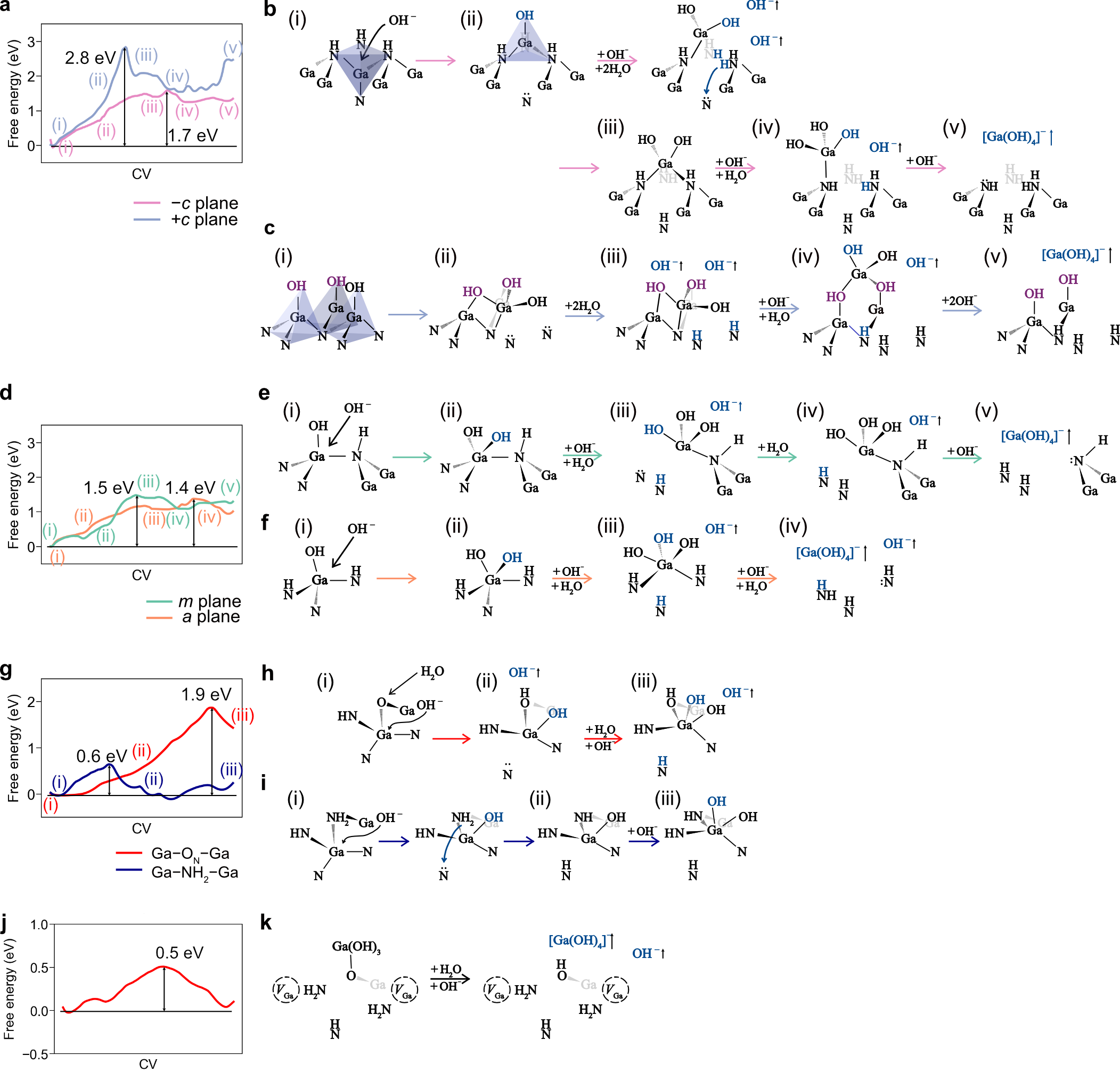}
    \caption{(a) Free energy surfaces comparing Ga dissolution pathways on polar surfaces ($-c$ and $+c$ planes). The corresponding structural evolutions are shown in (b) for the $-c$ plane and in (c) for the $+c$ plane. (d) Free energy surfaces comparing Ga dissolution pathways on polar surfaces ($m$ and $a$ planes). The corresponding structural evolutions are shown in (e) for the $m$ plane and in (f) for the $a$ plane. (g) Comparison of free energy surfaces for \ce{Ga-N} bond breaking in the presence of \ce{Ga-$\rm O_N$-Ga} and \ce{Ga-$\rm NH_2$-Ga} motifs. The corresponding pathways are shown in (h) and (i), respectively. (j) Free energy surface corresponding to the removal of a \ce{Ga-$\rm O_N$-Ga} bridge, with (k) depicting the associated structural change. In the structural models, reaction products and adsorbates are highlighted in blue, and atoms occluded by others are rendered with reduced opacity. In (c), pre-adsorbed \ce{OH-} on Ga ions that assists in breaking the \ce{Ga-N} bond are highlighted in purple. }
    \label{fig:pathway_total}
\end{figure*}

Figure~\ref{fig:pathway_total}a shows free-energy diagrams of the Ga dissolution pathways on the polar surfaces. On the $-c$ surface, Ga ions lie below \ce{-NH} units and have four \ce{Ga-N} bonds, all of which must be broken for Ga dissolution to occur. The absence of exposed \ce{Ga-OH} species, which could otherwise impede the approach of aqueous \ce{OH-} ions to the surface, allows a underlying Ga ion to form a \ce{Ga-OH} bond (\rom{1}$\rightarrow$\rom{2} in Figure~\ref{fig:pathway_total}b), with a moderate reaction energy barrier of $\sim$0.8 eV. This process results in an upward shift of the Ga ion, breaking one \ce{Ga-N} bond. The subsequent step for breaking another \ce{Ga-N} bond involves the \ce{OH-} adsorption and the \ce{H+} passivation of electron lone-pairs on two N ions (\rom{2}$\rightarrow$\rom{3}). It should be noted that, during this step, a proton transfer from the \ce{NH2} unit to the lower-lying N ion plays an important role in the \ce{H+} passivation. This step leads to an energy barrier of $\sim$0.8 eV. On the other hand, the breaking of the remaining two \ce{Ga-N} bonds (\rom{3}$\rightarrow$\rom{4} and \rom{4}$\rightarrow$\rom{5}) and the formation of \ce{[Ga(OH)4]-} proceeds with much smaller energy barriers below $\sim$0.2 eV. 

Unlike the $-c$ surface, the initial adsorption of aqueous \ce{OH-} to Ga ions on the $+c$ surface is expected to be hindered by Coulomb repulsion between an incoming \ce{OH-} ion and pre-adsorbed \ce{OH-} species. Indeed, our OPES simulations reveals an alternative pathway for Ga dissolution that is not initiated by the adsorption of aqueous \ce{OH-}. Specifically, the initial upward shift of a Ga ion occurs through the formation of \ce{Ga-OH-Ga} bonds, involving preexisting \ce{OH-} ions bound to neighboring Ga ions (\rom{1}$\rightarrow$\rom{2} in Figure~\ref{fig:pathway_total}c). This is followed by protonation of the exposed lone-pair states on adjacent N ions (\rom{2}$\rightarrow$\rom{3}). However, these steps cause a substantial energy barrier of $\sim$2.8 eV, because surface Ga ions are tightly bound to three underlying N ions, significantly restricting their upward displacement. Compared to the first two steps, the subsequent reactions, breaking a single \ce{Ga-N} bond (\rom{3}$\rightarrow$\rom{4}) and two \ce{Ga-OH} bonds (\rom{4}$\rightarrow$\rom{5}), proceed with relatively small energy barriers. 

Overall, our results highlight the relatively high etchability of the $-c$ surface in alkaline environments under typical experimental conditions, whereas the $+c$ surface exhibits significant resistance to etching—consistent with experimental observations.~\cite{review_ZHUANG20051,selective_etching_of_polar_surface} 

\subsubsection{Ga dissolution on nonpolar surfaces}

Figure~\ref{fig:pathway_total}d illustrates the free-energy profiles for Ga dissolution pathways on the non-polar surfaces. On the $m$ plane, the initial \ce{OH-} adsorption (\rom{1}$\rightarrow$\rom{2} in Figure~\ref{fig:pathway_total}e) is feasible, requiring a small energy barrier of $\sim$0.3 eV. However, this process is insufficient to break  \ce{Ga-N} bonds. Additional \ce{OH-} adsorption leads to the breaking of two \ce{Ga-N} bonds (\rom{2}$\rightarrow$\rom{3}), with an energy barrier of approximately 1.2 eV. During this step, one of the resulting N lone-pair states is immediately passivated by a proton. The proton passivation of the remaining lone-pair state occurs next (\rom{3}$\rightarrow$\rom{4}), following a benign reaction pathway. Further adsorption of \ce{OH-} onto the dissolving Ga ion leads to its dissociation from the surface (\rom{4}$\rightarrow$\rom{5}), with a small energy barrier of $\sim$0.2 eV.

On the $a$ plane, the initial two steps of Ga dissolution are analogous to those on the 
$m$ plane: two consecutive \ce{OH-} adsorption events expose a lone-pair state on a N ion by breaking a \ce{Ga-N} bond, which is passivated by a proton (\rom{1}$\rightarrow$\rom{2} and \rom{2}$\rightarrow$\rom{3} in Figure~\ref{fig:pathway_total}f). However, in this case, the energy barrier associated with the bond breaking is relatively lower than that on the $m$ plane, as only a single \ce{Ga-N} bond is broken during this stage. Instead, the $a$ plane exhibits a higher energy barrier ($\sim$0.4 eV) in the subsequent \ce{[Ga(OH)4]-} desorption step (\rom{3}$\rightarrow$\rom{4}), which involves additional \ce{OH-} adsorption and immediate passivation of the resulting N lone-pair state.


The effective activation energy of the pathways, corresponding to the difference between the highest and lowest energies, is found to be 1.5 eV for the $m$ plane and 1.4 eV for the $a$ plane. These values are comparable to that of the $-c$ plane, confirming the etchability of these non-polar surfaces in alkaline solutions. In addition, experimental results show that wet etching of the $m$ plane is slightly slower than, or comparable to, that of the $a$ plane,\cite{Nanorod_RYU2024160040} consistent with the comparable activation energies predicted by our simulations.

\subsubsection{Role of oxygen bridges}

As we demonstrated above, oxygen bridges form intermittently during the etching process. In the case of oxygen bridges involving O$_{\rm{int}}$, the brides can be dissociated through reactions with \ce{H2O} and \ce{OH-}, effectively resulting in a structure equivalent to that formed by the adsorption of two \ce{OH-} ions (Figure S10). According to our MD simulations, this process readily occurs. 

Conversely, oxygen bridges in which an O ion occupies a N site can remain persist for a longer duration. Figure~\ref{fig:pathway_total}g presents the free-energy profile for the initial steps of Ga dissolution within a \ce{Ga-O$_{\rm{N}}$-Ga} bridge on the $m$ plane. Interestingly, the \ce{Ga-O-Ga} bonds are maintained even when a bridged Ga ion is attacked by \ce{OH-}; instead, the reaction favors breaking a \ce{Ga-N} bond. This leads to the formation of a lone-pair electron state on a N ion, which is subsequently passivated by a proton (see Figure~\ref{fig:pathway_total}h). This pathway exhibits a substantial activation energy of approximately 1.9 eV, indicating slow reaction kinetics. In contrast, Ga ions that lose neighboring N ions without forming oxygen bridges can readily coordinate with \ce{OH-} (Figure~\ref{fig:pathway_total}i). This pathway results in a smaller energy barrier of $\sim$0.6 eV and is thus expected to proceed rapidly. 

In light of the results presented in Figure~\ref{fig:pathway_total}g, the dissolution of Ga ions around a \ce{Ga-O$_{\rm{N}}$-Ga} bridge is likely to occur before \ce{OH-} adsorption onto the bridged Ga ions. During the dissolution of such neighboring Ga ions, N ions connected to \ce{Ga-O-Ga} bridges can acquire protons, thereby weakening their chemical bonds with the bridged Ga ions. As a result, subsequent \ce{OH-} adsorption onto the bridged Ga ions becomes more feasible, leading to \ce{Ga-O-Ga} configurations that are only weakly connected to the surface, as illustrated in 
Figure~\ref{fig:pathway_total}k. These configurations are indeed frequently found in our etching simulations. The following generation of \ce{[Ga(OH)4]-} can efficiently proceed with a low activation energy of approximately 0.5 eV (see Figure~\ref{fig:pathway_total}j), eliminating the oxygen bridges. Note that the conclusions drawn from this analysis on the $m$ plane are also applicable to the other surfaces, considering the results of the corresponding etching simulations.

As noted in Section 2.2, the concentration of oxygen bridges formed during wet etching is not significant, suggesting a limited impact on the overall etching rate. However, since the removal of each oxygen bridge is delayed until the dissolution of neighboring Ga and N ions, these configurations are likely to be present on the surface after alkaline etching. Notably, previous studies have demonstrated that substitutional oxygen in GaN can form critical defect complexes that enhance non-radiative recombination of charge carriers.\cite{VGa-ON_epxt_PhysRevB.80.153202,VGa-ON_calc_10.1063/1.4942674} In this context, the role of \ce{Ga-O$_{\rm{N}}$-Ga} bridges, likely formed on the sidewalls of GaN nanorods during wet etching, requires further investigation through both experimental and theoretical studies.


\section{Conclusion}

We presented a comprehensive atomic-level investigation of GaN wet etching in KOH solution using large-scale NNP-MD simulations. By using an iterative learning strategy, we developed a Behler–Parrinello-type NNP with a high capability to describe chemical reactions associated with the alkaline etching of GaN surfaces. We showed that the NNP-MD simulations accurately reproduce the structural modification of GaN nanorods observed in experiments. The etching simulations revealed the morphologies of etched surfaces: pyramidal pits form on the $-c$ surface, while truncated pyramidal pits develop on the $+c$ surface. The non-polar ($a$ and $m$) surfaces exhibit highly anisotropic lateral etch propagation, resulting in planar etched morphologies. Key surface reactions involved in etching were identified through atomic trajectory analysis, and OPES simulations provided free-energy profiles for Ga dissolution, a process critical to the etching kinetics. The moderate activation barriers observed on the $-c$, $a$, and $m$ planes indicate their high etchability, whereas the significantly higher barrier on the $+c$ plane accounts for its etch resistance. Additionally, we showed that \ce{Ga-O-Ga} bridges can be present on etched surfaces, which may deteriorate the optoelectronic performance of GaN-based devices. The detailed insights from our study advance the fundamental understanding of GaN surface chemistry during alkaline etching and support the rational design of surface processes in nitride-based device fabrication.

\section{Experimental Section}

\threesubsection{Neural network potential}
We build a Behler-Parrinello-type NNP trained with the \texttt{SIMPLE-NN} package.~\cite{BPNN_PhysRevLett.98.146401,simple-nn_LEE201995} Atom-centered symmetry functions (ACSFs) are used as input features, with a cutoff radius of 6 \AA\ for Ga, N, K, and O and 4.5 \AA\ for H. The feature vector for each element initially contains 310 components. To enhance the training and inference efficiency of the NNP, we reduce the size of the feature vectors by applying CUR decomposition (see Section S1 Supporting Information for details). As a result, the final feature vector sizes become 145, 141, 51, 153, and 151 components for Ga, N, K, H, and O, respectively. Specific parameters for ACSFs are summarized in Table S1 in Supporting Information.

The feature vectors are scaled using the maximum and minimum values. To speed up the learning process, we apply principal component analysis to the feature vectors and whitening them. We employ a fully-connected neural network architecture consisting of two 30-30 hidden layers. The dataset is split into 90\% for training and 10\% for validation. Training is conducted in two stages: an initial stage for generating a baseline model and an iterative stage for model refinement.~\cite{HCH_HF_etching} In the initial stage, we use a learning rate of $10^{-4}$ and a batch size of 4. In the iterative stage, these parameters are adjusted to $10^{-5}$ and 8, respectively. The Adam optimizer is used for optimization.

\threesubsection{DFT calculations}

We perform DFT calculations using the Vienna Ab initio Simulation Package (\texttt{VASP}) with PAW pseudopotentials.~\cite{VASP_KRESSE199615} The Perdew-Burke-Ernzerhof (PBE) functional is used to approximate the exchange-correlation energy between electrons.~\cite{GGA_PhysRevLett.77.3865} The semicore $d$ states for Ga and $p$ states for K are treated as valence states. A energy cutoff for plane wave basis is set to 450 eV. We sample only the $\Gamma$ point for Brillouin zone integration because the supercell size for generating the training data is sufficiently large. We account for Van der Waals interactions with the Grimme-D3 method.~\cite{D3} During AIMD simulations, a timestep is set to 1 fs when hydrogen, the lightest element, is present in the supercell for ensuring the stability of the simulations. Otherwise, the timestep is set to 2 fs. Temperature control is implemented using the Nosé-Hoover thermostat for NVT simulations and the Langevin thermostat for NPT simulations.

\threesubsection{MD simulation for wet etching in KOH solution}
We perform molecular dynamics simulations of wet etching in KOH solution using the NNP within the \texttt{LAMMPS} package.~\cite{LAMMPS} To efficiently explore the evolution of surface morphology within accessible MD time scales, we employ the TAD method, which accelerates chemical reactions by elevating the temperature. Specifically, we conduct NPT simulations at 2000 K, a temperature below the melting point of GaN. To prevent water vaporization, a pressure of 100 kbar is applied. A time step is set to 0.5 fs to ensure the stability of MD simulations. Due to the limited supercell size, the concentrations of dissolved species, such as gallium hydroxide ions ([Ga(OH)$_{4}$]$^{-}$) and ammonia (\ce{NH_3}), can instantly become unrealistically high, potentially altering solution properties and promoting undesirable reactions among byproducts. To mitigate this, we regularly monitor the amount of dissolved species during etching simulations and remove them as necessary. At the same time, we replenish water and hydroxide molecules to maintain charge neutrality and to preserve the pH condition. For instance, when a [Ga(OH)$_{4}$]$^{-}$ ion is removed, one OH$^-$ and two H$_2$O molecules are added. When a NH$_3$ molecule is removed, one H$_2$O molecule is added. 

\threesubsection{OPES (On-the-fly probability enhanced sampling)}
To determine the free-energy profiles of key chemical reactions under experimental etching conditions, we conduct enhanced-sampling simulations based on collective variables (CVs). We define CVs as the coordination numbers of Ga–N, Ga–O, and N–H bonds. Herein, we adopt a continuous function to describe the coordination number ($ \mathrm{CN}^{A}_{B}$) which quantifies the number of neighboring atoms of type $A$ around a central atom $B$ within a cut-off radius $r_0$: 
 \begin{equation}
    \label{equation:CN}
    \mathrm{CN}^A_B = \sum_{i\in A}\frac{1-(d_{i}/r_0)^l}{1-(d_{i}/r_0)^m},
 \end{equation}
where $l$ and $m$ are exponents controlling the sharpness of the function and $d_{i}$ is the distance between atom $i$ and the central atom $B$. The parameters $l$, $m$, and $r_0$ are tuned for each bond type: Ga$-$N (14, 30, and 2.47 \AA), Ga$-$O (12, 30, and 2.80 \AA), and N$-$H (16, 30, and 1.58 \AA). To enhance the likelihood of identifying transition states and the accuracy of free-energy estimation, we employ an adaptive reaction coordinate, $\sigma(\mathbf{z})$, where $\mathbf{z}$ is a set of collective variables $\{\rm{CN^N_{Ga},CN^O_{Ga},CN^H_N}\}$. The coordinate $\sigma(\mathbf{z})$ evolves along a parameterized curve, $s(\sigma)$, which represents the average transition path connecting two local minima. Detailed information on this approach and its implementation can be found in previous literature.\cite{PathCV_PhysRevLett.109.020601,path_doi:10.1021/acs.jctc.3c00938}   

To search for the lowest free-energy pathways, we employ the on-the-fly probability enhanced sampling (OPES) method, which significantly improves the convergence of the calculations.\cite{OPES_doi:10.1021/acs.jpclett.0c00497} In OPES, the equilibrium probability distribution is estimated on the fly, followed by constructing a bias potential to guide the system toward a desired target distribution. A well-tempered target distribution, characterized by a bias factor $\gamma > 1$ and a temperature-dependent parameter $\beta=1/k_BT$, is considered in the present study. The bias potential $V_n(\sigma)$, applied to the reaction path at each iteration, is given as:
\begin{equation}
    \label{equation:OPES}
    V_n(\sigma) = (1-1/\gamma)\frac{1}{\beta} \log \left(\frac{P_n(\sigma)}{Z_m}+\epsilon \right), 
\end{equation}
with the probability distribution at $n$-th iteration $P_n(\sigma)$, a normalization factor $Z_n$, and a regularization parameter $\epsilon=e^{-\beta\Delta E/(1-1/\gamma)}$. The maximum value of the bias potential is set to 3 eV to prevent the potential from overflowing into undesired high-energy molecular configurations. We employ the \texttt{plumed2} package to conduct OPES simulations.\cite{PLUMED2_TRIBELLO2014604} 


\medskip
\textbf{Supporting Information} \par 
Supporting Information is available from the Wiley Online Library or from the author.

\medskip
\textbf{Acknowledgements} \par 
This work was supported by Samsung Electronics Co., Ltd(IO201214-08143-01) and the Nano \& Material Technology Development Program through the National Research Foundation of Korea (NRF) funded by Ministry of Science and ICT (RS-2024-00407995)

\medskip

%
\bibliographystyle{MSP}
\bibliography{main}






\begin{figure}
\textbf{Table of Contents}\\
\medskip
  \includegraphics{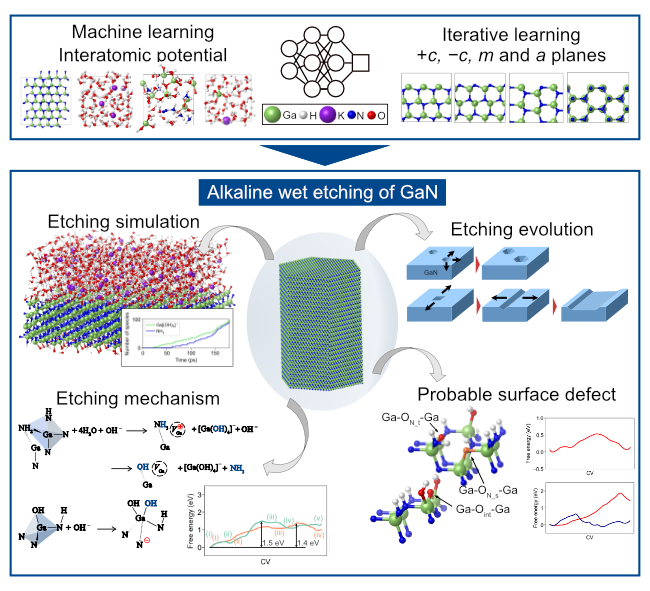}
  \medskip
  \caption*{ToC Entry}
\end{figure}

\end{document}